# Determination of oxygen vacancy limit in Mn substituted YSZ ceramics

Joanna Stępień[a,*], Marcin Sikora[b], Czesław Kapusta[a], Daria Pomykalska[c], Mirosław M. Bućko[c]

[a]*AGH University of Science & Technology, Faculty of Physics & Applied Computer Science, Department of Solid State Physics, Mickiewicza 30, 30-059 Krakow, Poland*
[b]*AGH University of Science and Technology, Academic Centre for Materials and Nanotechnology, Mickiewicza 30, 30-059, Kraków, Poland*
[c]*AGH University of Science & Technology, Faculty of Materials Science & Ceramics, Department of Ceramics and Refractory Materials, Mickiewicza 30, 30-059 Krakow, Poland*

**Abstract:**

The series of $Mn_x(Y_{0.148}Zr_{0.852})_{1-x}O_{2-\delta}$ ceramics was systematically studied by means of X-ray absorption and emission spectroscopy and DC magnetic susceptibility. XAS and XES results show changes in manganese oxidation state and a gradual evolution of the local atomic environment around Mn ions upon increasing dopants content, which is due to structural relaxation caused by growing amount of oxygen vacancies. Magnetic susceptibility measurements reveal that $Mn_3O_4$ precipitates are formed for $x \geq 0.1$ and enable independent determination of actual quantity of Mn ions dissolved in YSZ solid solution. We show that amount of oxygen vacancies generated by manganese doping into YSZ is limited to 0.19 per formula unit.

## 1. Introduction

Yttria Stabilized Zirconia (YSZ) is commonly used as electrolyte in Solid Oxide Fuel Cells (SOFCs). This type of fuel cell operates at high temperatures (~1000°C), which enables oxygen ions conductivity, but is detrimental to materials used in construction limiting power output [1]. During operation SOFC undergoes gradual degradation due to destructive interactions at interfaces and surfaces of the materials building the cell [2, 3]. Among the possible ways to ameliorate cell operation is intentional mixing of electrodes and electrolyte materials in order to improve theirs compatibility and stability [4, 5]. In particular the application of MIECs (Mixed Ionic-Electronic Conductors) as a cathode material [6] or as thin layer on the surface of the electrolyte [7]. Therefore manganese doping into cubic YSZ was proposed for modification of typical SOFC electrolyte working in contact with $LaMnO_3$ cathode [1, 8]. During cell operation Mn diffuses into YSZ (both materials interact with each other) [9], which causes decrease in lattice constant and changes in local atomic structure around dopant ions [10], stabilizes YSZ cubic structure [11, 12] or induces partial transformation to the tetragonal phase [13]. Above the solubility limit Mn, that is approx. 8 − 15 mol% (depending on synthesis conditions) [14] can form $Mn_2O_3$ 7 or $Mn_3O_4$ [15] precipitates.

---

[*] Corresponding author
  e-mail address: jstepien@agh.edu.pl, tel. (+48)12 6175254, fax. (+48)12 6341247

Important parameter that influences ionic conductivity and thus cell performance is the number of oxygen vacancies generated by aliovalent dopants in $ZrO_2$. Yttrium (3+) substituting for zirconium (4+) generates oxygen vacancies necessary to stabilize $ZrO_2$ in cubic structure. Manganese introduces additional vacancies, but as a dopant it can adopt different oxidation states and thus the structural and chemical changes it introduces with respect to doped material are difficult to predict. To our knowledge the oxidation state of manganese ions doped into YSZ has not been reported yet, thus it is uncertain how much vacancies can be introduced in this way.

We present the study of cubic YSZ doped with different concentrations of manganese for the purpose of determining local structural properties, especially those seen from Mn site. We have employed X-ray spectroscopic techniques, namely K-edge XANES (X-Ray Absorption Near Edge Structure) and Kβ XES (X-Ray Emission Spectroscopy) [16,17], which enable selective probing of elemental sample constituents. Spectral shape of XANES is influenced by local atomic structure. It provides information about the electronic state, particularly oxidation state of selected element. Kβ XES spectrum depends on exchange interaction between localized 3p electronic orbitals and unpaired d-electrons of the valence band and in result gives information about effective spin independent of changes in the local atomic structure. DC susceptibility probes the bulk magnetic properties of the material. When performed in a wide temperature range it allows to identify type and quantify amount of magnetic precipitates. The objective of this work was to determine oxidation state of Mn ions and thus to quantify the amount of oxygen vacancies in YSZ ceramics generated by dopants.

## 2. Materials and methods

Stabilized cubic zirconia solid solutions containing manganese $Mn_x(Y_{0.148}Zr_{0.852})_{1-x}O_{2-\delta}$, with nominal content of Mn x = 0; 0.025; 0.05; 0.1; 0.15; 0.2 and 0.25 mol% were prepared by means of co-precipitation-calcination method and then sintered for 2h at 1500°C. According to XRD characterization, the two samples with the highest concentration of Mn contain small amounts of $Mn_3O_4$ phase (3.6 and 7 wt% for x = 0.2 and 0.25, respectively) [18].

XANES measurements were performed on beamline C of Doris III synchrotron at HASYLAB, Hamburg, using fluorescence detection mode. Mn K-edge spectra were collected from $Mn_3O_4$ and $MnO_2$ references and all doped samples (x = 0.025 – 0.25). In addition, Y and Zr K edges were measured for the four selected samples (x = 0; 0.1; 0.15; 0.25) and pure $ZrO_2$ as a reference. Data were analyzed using ATHENA software (part of Demeter package) [19].

XES spectra were collected on XAS-XES ID26 beamline of ESRF, Grenoble. Mn $K\beta_{1,3}$ line was measured using bending crystal Si (440) analyzer on all doped samples and reference samples: MnO, $LaMnO_3$ and $CaMnO_3$. Energy of incident radiation was set to 6.7keV.

DC susceptibility measurements were performed using Vibrating Sample Magnetometry [20] (VSM) option of Quantum Design Physical Property Measurements System Model 6000. Magnetization M(T) curves were measured from 2 K to 300 K in steady magnetic fields of 0.1T and 2T.

# 3. Results

## 3.1. XANES

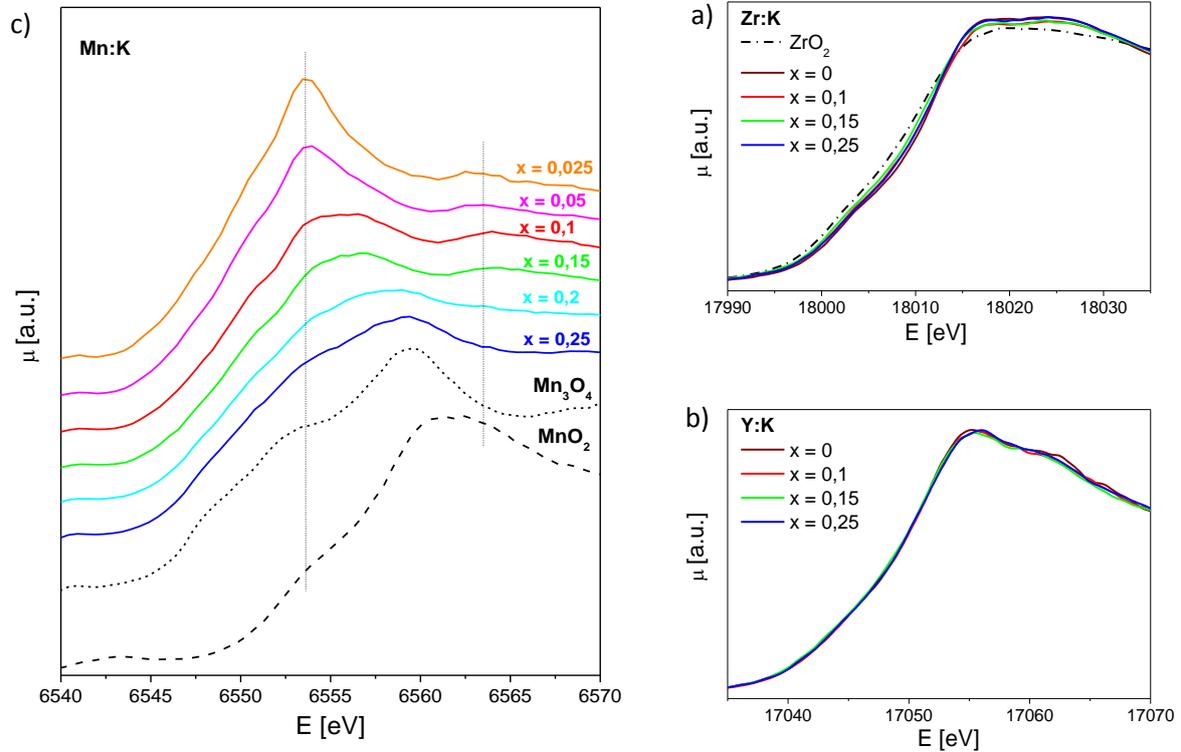

**Fig.1.** XANES spectra of a) Zr, b) Y, c) Mn K edges for $Mn_x(Y_{0.148}Zr_{0.852})_{1-x}O_{2-\delta}$. Vertical lines are guide for eyes to visualize changes of the white line and the next maximum above it.

**Table 1**
Mn K edge energy positions for $Mn_x(Y_{0.148}Zr_{0.852})_{1-x}O_{2-\delta}$ samples and references.

| x/sample | $E_0$ [eV] |
|---|---|
| 0.025 | 6547.26 |
| 0.05 | 6547.72 |
| 0.1 | 6548.48 |
| 0.15 | 6548.75 |
| 0.2 | 6548.79 |
| 0.25 | 6548.54 |
| $MnO_2$ | 6553.12 |
| $Mn_3O_4$ | 6548.43 |

XANES spectra at Mn, Zr and Y K absorption edges are shown in Fig.1. With increasing Mn content there are significant changes in Mn:K spectra but only minor changes in Zr:K spectra and practically no changes in Y:K spectra. This suggests that the biggest changes with increasing dopping happen around manganese ions. Zr:K edge of the samples (cubic structure) is also not much different from the edge of pure $ZrO_2$ (monoclinic structure). There is only slight change in edge position and height of the white line but the shape of $ZrO_2$ spectra is very similar to the samples. For Mn:K edge we see change in edge position and significant change in edge shape which implies changes in oxidation state, local environment and density of unoccupied electronic states around Mn ions. The edge position moves to higher energies with increase in doping which shows that the oxidation state of Mn ions increases with increasing Mn content [21, 22].

To determine exact oxidation state of Mn ions in each sample we have also measured reference spectra with known mean oxidation states and similar local environments and oxidation state for Mn ions: $MnO_2$ and $Mn_3O_4$ oxides. We can see that $Mn_3O_4$ spectrum is very similar in shape and edge position to two samples with the highest Mn content.

For reference samples we assume formal oxidation states according to ionic model, namely 4+ ($MnO_2$) and 2,67+ ($Mn_3O_4$). For the YSZ+Mn samples we determine the mean oxidation state based on the energy shift of the edge at half value of the absorption step. Assuming linear dependence between energy edge position and mean oxidation state, the latter was calculated in the doped samples by comparing edge energy in doped samples and reference samples. The uncertainties were estimated taking into account the accuracy of energy measurement in the experiment which was 0.1 eV. Calculated mean oxidation states for Mn ions in each sample increases from 2.34(3) for x = 0.025 to 2.77(3) for x = 0.2 and then decreases slightly to 2.70(3) for x = 0.25. On average Mn oxidation state estimated from XANES spectra is close to 2.5+, which suggests mixed valence character of the solid solution.

### 3.2. XES

$K\beta_{1,3}$ Mn lines of reference samples MnO (Spin = 1.5), $LaMnO_3$ (S = 2), $CaMnO_3$ (S = 2.5) and measured samples normalized to unit area are shown in Fig.2. top panel. Comparison with the reference samples shows that $K\beta$ emission lines are located close to that of $LaMnO_3$ line indicating mean Mn spin close to 2 in measured samples. With increasing Mn doping the $K\beta$ maximum shifts slightly to lower emission energy. In order to enhance and better visualize the evolution within the series the same spectra are shown upon subtraction of $LaMnO_3$ spectrum in the bottom panel of Fig.2. Differential spectrum of x = 0.025 sample shows negative feature at 6.4897 eV and positive peak at 6.4915 eV. With increasing Mn content the feature at 6.4897 eV decreases its negative intensity and finally changes to positive value. The peak at 6.4915 eV also decreases but remains positive for all the samples studied. This indicates that there exist at least two components forming the spectra, which we attribute to separate phases containing manganese ions, for instance YSZ+Mn solid solution and Mn$_3$O$_4$ precipitates.

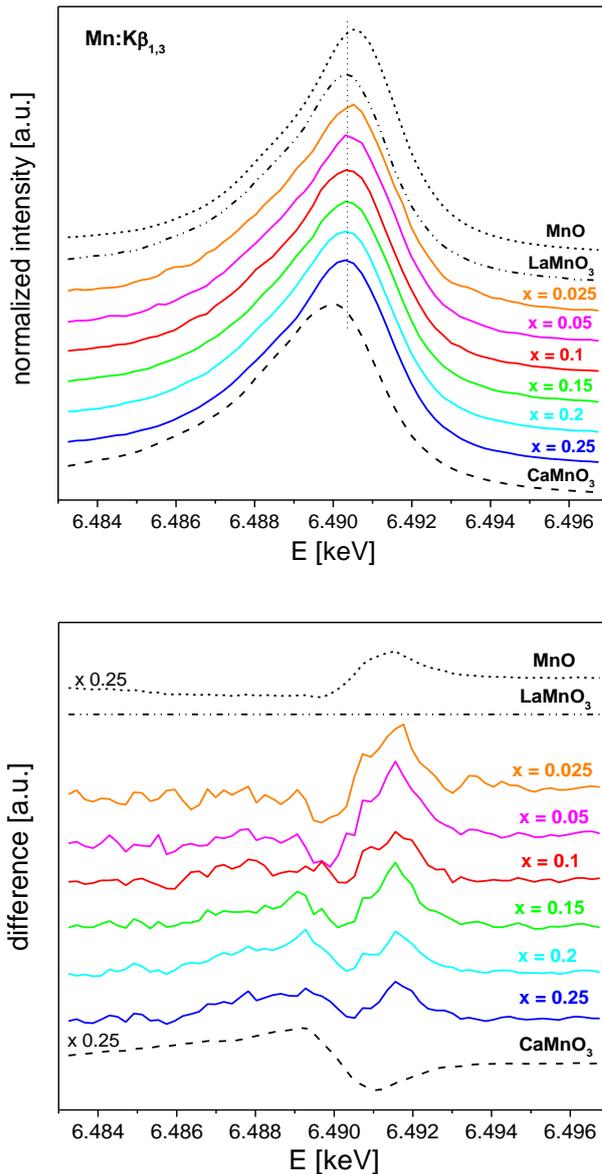

**Fig.2.** $K\beta_{1,3}$ Mn lines of the samples $Mn_x(Y_{0.148}Zr_{0.852})_{1-x}O_{2-\delta}$ and reference samples MnO, $LaMnO_3$ and $CaMnO_3$ normalized to unit area (top panel). Differential spectra of the samples and the reference samples with subtracted $LaMnO_3$ spectrum (bottom panel). Vertical line in the top panel is a guide for eyes to visualize changes of the main peak position.

**Table 2**
The positions of the Mn K$\beta_{1,3}$ lines of Mn$_x$(Y$_{0.148}$Zr$_{0.852}$)$_{1-x}$O$_{2-\delta}$ series and reference compounds determined using first moment analysis. The average spin value of Mn dopants is estimated from linear dependence between K$\beta_{1,3}$ position and the nominal spin of reference samples following ref. [16].

| x/sample | $\overline{E_{K\beta}} - 6489\ [eV]$ | spin |
|---|---|---|
| 0.025 | 0.541 | 2.08(2) |
| 0.05 | 0.542 | 2.08(2) |
| 0.1 | 0.487 | 2.00(2) |
| 0.15 | 0.480 | 1.99(2) |
| 0.2 | 0.463 | 1.96(2) |
| 0.25 | 0.466 | 1.97(2) |
| MnO | 0.815 | 1.5 |
| LaMnO$_3$ | 0.459 | 2 |
| CaMnO$_3$ | 0.129 | 2.5 |

To estimate effective Mn ions spin quantitatively we calculated the first moment of measured K$\beta$ XES spectra on which spin is linearly dependent with good accuracy [16]. Mn ions spin of the samples is calculated assuming formal spin of the Mn in reference samples by comparison of the first momenta of the K$\beta$ main spectra. The results are shown in Table 2. Taking into the account the uncertainties of this estimation we can divide the samples into two groups: for two samples with the smallest Mn content, i.e. without Mn3O4 precipitates, the value of the spin is virtually identical and significantly higher than in the other samples. The spin value higher than 2 suggest large amount of Mn$^{2+}$ ions diluted in YSZ.

### 3.3. DC susceptibility

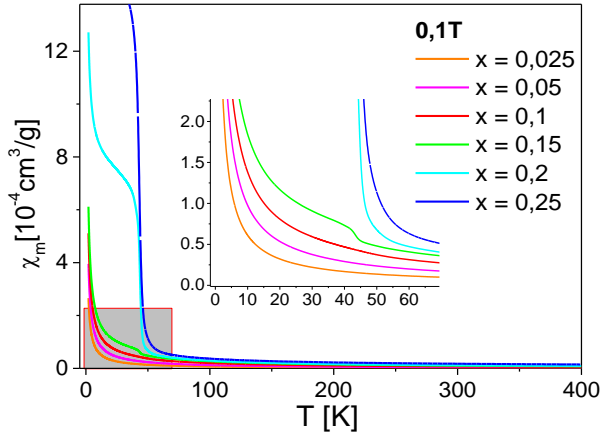
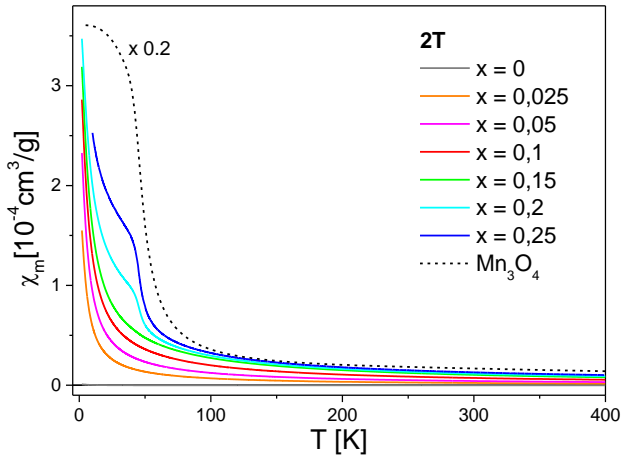

**Fig.3.** Mass magnetic susceptibility at 0.1T (top panel) and 2T (bottom panel) for Mn$_x$(Y$_{0.148}$Zr$_{0.852}$)$_{1-x}$O$_{2-\delta}$ samples. At the bottom panel the data for ZrO$_2$ and Mn$_3$O$_4$ are also shown.

Mass magnetic susceptibility curves derived from magnetization M(T) measurements for all the samples studied and the reference sample Mn$_3$O$_4$ at 0.1Tesla and 2Tesla of the applied magnetic field induction are shown in Fig.3. For the samples with x < 0.15 a typical Curie-like hyperbolic temperature dependence is observed. For the samples with x ≥ 0.15 a step on the top of hyperbolic temperature dependence is observed at about 50K, gradually increasing with growing Mn content. Comparing these features with Mn$_3$O$_4$ reference sample we observe the same behavior of susceptibility at this temperature. The magnitude of the step can be considered as a quantitative indication of the $Mn_3O_4$ phase precipitation. Thus from our magnetic measurements we can see that already the sample with Mn content x = 0.15 reveals some amount of $Mn_3O_4$ precipitations, which is below the limit that could be detected by X-ray diffraction measurements [18]. The pristine $ZrO_2$ sample exhibits magnetic susceptibility which is more than two orders of magnitude lower (horizontal line close to zero in the plot) and negative (diamagnetic) except for low temperatures where it changes its sign to positive, indicating appearance of a weak paramagnetic contribution.

## 4. Discussion

The valence state of ions that substitute Zr (4+) in $ZrO_2$ affects the amount of oxygen vacancies generated in the solid solution. Number of vacancies is one of the crucial parameters determining ionic conductivity of solid electrolyte. Assuming formal oxidation state of Y (3+) ions, they introduce 1/2·0.148(1-x) oxygen vacancies per formula unit in $Mn_x(Y_{0.148}Zr_{0.852})_{1-x}O_{2-\delta}$ solid solution. Manganese ions of unknown mean oxidation state V generate x(2-V/2) vacancies per formula unit. Our goal was to determine the mean oxidation state of Mn ions from X-ray spectra assuming that Mn is in the high spin state. Taking into account that the samples with high manganese content are composed of $Mn_3O_4$ phase in addition to YSZ+Mn solid solution phase, we consider only Mn ions dissolved in YSZ as a creators of oxygen vacancies. We calculated the amount of $Mn_3O_4$ phase in x = 0.2 and

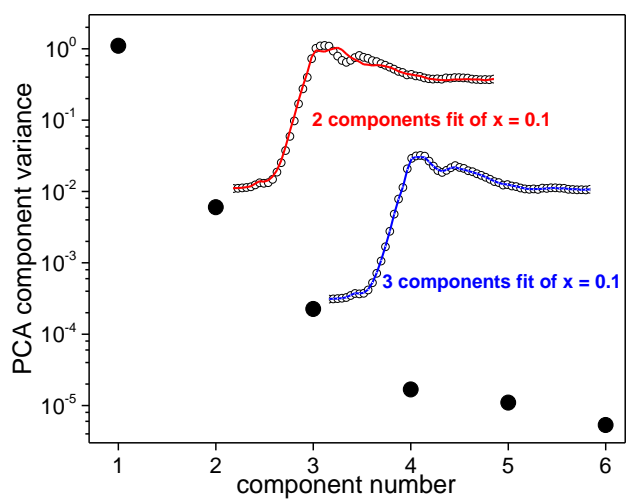

**Fig.4.** Variances of components from PCA analysis of Mn K-edge XANES spectra of Mn doped YSZ. Insets present linear combination fitting of spectrum of x = 0.1 with two (x = 0.025 & 0.25 spectra) and three (x = 0.025, 0.15 & 0.25 spectra) components. Both methods confirms that all the spectra of the series can be satisfactory reproduced using at least three independent components that are attributed to three different local structure environments (phases) involving Mn dopants (details in text).

x = 0.25 to be 2wt% and 4.4wt% respectively, that results in 14.8% and 25.6% of all Mn ions precipitating into $Mn_3O_4$ phase. In the case of x = 0.1 and x = 0.15 the precipitations are negligible (below 1% of all Mn ions) and do not influence further analysis. Amount of precipitates was also determined by Pomykalska et. al. from X-Ray Diffraction measurements. They determined the amount of $Mn_3O_4$ phase in x = 0.2 and x = 0.25 to be 3.6wt% and 7wt%, respectively, that results in 26.7% and 40.7% of Mn ions in the $Mn_3O_4$ phase, respectively. Discrepancy between these two method is attributed to different volume of the materials probed. The XRD pattern is formed by significantly large ordered volumes of the sample, while magnetic susceptibility probes predominantly the effective magnetic moment of Mn dopants irrespective of crystal structure. Moreover, the numbers estimated by Pomykalska et al. were obtained under assumption of sharp solubility limit of x = 0.18, which according to our XANES results is not entirely correct. The Mn solubility decreases gradually, but is not sharply limited as indicated by principal component analysis (PCA) and linear combination fits of XANES spectra. PCA analysis [23] indicates that minimum three independent components are needed to fit all datasets. Also linear combination fit of XANES spectra with two standards, i.e. $x = 0.25$ sample (as precipitate rich) and $x = 0.025$ sample (as pure YSZ phase with dissolved Mn ions), is poor, while addition of third standard results in reasonable fit (Fig.4.). As such we show that changes occurring in the samples with increasing doping cannot be explained solely by precipitation of $Mn_3O_4$ phase. The real picture is more complex – local structure around Mn dopands in the YSZ+Mn solid solution evolves gradually due to increasing number of oxygen vacancies. Tendency to precipitation of $Mn_3O_4$ is observed already at lower Mn concentration and only the volume of precipitates grows with doping level. Therefore, straightforward estimation of mean oxidation state using solely Mn K-edge shift of XANES spectra is doubtful due to evolution of local symmetry of absorber. Kβ XES spectra, that are significantly less sensitive to local structure, seem to be better indicator of electronic population of dopants.

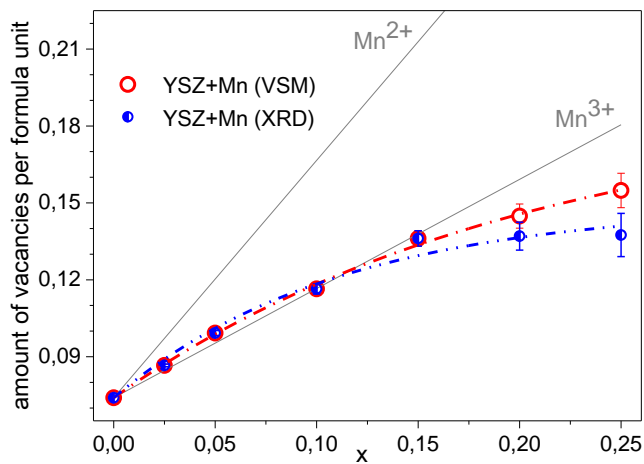

Fig.5. Oxygen vacancy concentration dependence on Mn content calculated from XES measurements for the YSZ+Mn solid solution phase of the samples. The amount of Mn ions in solid solution was determined from VSM (red empty circles) and XRD (blue half-filled circles) measurements. The dashed lines are the exponential fits of presented dependencies.

Assuming high spin configuration and Mn spin in $Mn_3O_4$ phase (S=2.17), the spin value of Mn dissolved into YSZ can be estimated from weighted average and further transformed to mean oxidation state and expected amount of oxygen vacancies. (Fig.5) It is clear that amount of oxygen vacancies shows asymptotic behavior, that on one hand is inconsistent with the model of sharp solubility limit, and on the other hand allows to estimate the maximum amount of oxygen vacancies created by Mn doping in YSZ based on exponential fit. It equals to 0.190(8) and to 0.149(7) for estimation of the amount of $Mn_3O_4$ precipitate from magnetic susceptibility and XRD measurements, respectively.

## 5. Conclusions

Mean oxidation state of manganese ions increases with doping along with gradual evolution of local atomic structure around dopants. Mean oxidation state and effective spin of Mn ions were determined from XANES and XES spectra, respectively.

Magnetic susceptibility measurements reveal the presence of $Mn_3O_4$ phase already for $x = 0.15$ sample. They were used to quantify the amount of manganese dissolved in YSZ+Mn solid solution, which is significant for $x = 0.2$ and 0.25 and amounts to 2.0 and 4.4 wt%, respectively.

The amount of oxygen vacancies in the Mn+YSZ solid solution phase generated by doping increases gradually, however, it saturates due to Mn valence change and $Mn_3O_4$ precipitation. The maximum amount of oxygen vacancies generated by Mn doping of YSZ phase was estimated to 0.190(8) and to 0.149(7) per formula unit, based on estimation of the amount of $Mn_3O_4$ precipitate derived from magnetic susceptibility and XRD measurements, respectively.

**Acknowledgements**

Access to ESRF was supported by special project ESRF/73/2006 from the Polish Ministry of Science and High Education. JS and MS acknowledge support from the National Science Centre of Poland.